\documentclass[%
 reprint,
 amsmath,amssymb,
 prl,
]{revtex4-1}
\usepackage{graphicx}
\usepackage{dcolumn}
\usepackage{bm}

\begin{document}
\preprint{APS/123-QED}

\title{Non-stationary resonance dynamics of weakly coupled pendula}

\author{Leonid I. Manevitch}
\affiliation{%
 Institute of Chemical Physics, Russian Academy of Science.\\
}%


\author{Francesco Romeo}
\affiliation{
 Department of Structural and Geotechnical Engineering, SAPIENZA University of Rome.\\
}%
%
%
\date{\today}

\begin{abstract}
In this paper we fill the gap in understanding the non-stationary resonance dynamics of the weakly coupled pendula model, having significant applications in numerous fields of physics such as superconducting Josephson junctions, Bose-Einstein condensates, DNA, etc.. While common knowledge of the problem is based on two alternative limiting asymptotics, namely the quasi-linear approach and the approximation of independent pendula, we present a unified description in the framework of new concept of Limiting Phase Trajectories (LPT), without any restriction on the amplitudes of oscillation. As a result the conditions of intense energy exchange between the pendula and transition to energy localization are revealed in all possible diapason of initial conditions. By doing so, the roots and the domain of chaotic behavior  are clarified as they are associated with this transition while simultaneously approaching the pendulum separatrix. The analytical findings are corroborated by numerical simulations. By considering the simplest case of two weakly coupled pendula, we pave the ground for new opening possibilities of significant extensions in both fundamental and applied directions.
\end{abstract}

\pacs{Valid PACS appear here}
\maketitle


The model of coupled pendula and some of its modifications play a significant role in Mechanics \cite{Scott}, Solid State Physics \cite{Braun} including superconducting Josephson junctions \cite{Likharev, Hadley, Braun2}, Photonics, including Bose-Einstein condensates \cite{Catal}, Biophysics, including DNA functioning \cite{Yaku}. The majority of the results in all these fields relate to stationary dynamics and are based on the fundamental stationary regimes, namely the Nonlinear Normal Modes (NNMs) in finite systems \cite{Man1, Vak1} and solitons (breathers) in infinite models \cite{Scott, Braun}. As for non-stationary processes, NNMs can also be used for their description provided that the intermodal resonance is absent \cite{Nayfeh}. However, the non-stationary resonance dynamics of finite systems turns out to be much more complicated. Therefore only isolated and predominantly numerical results were obtained in this field \cite{Sepul, Man_A}. The recently developed concept of Limiting Phase Trajectories (LPTs) allowed for a systematic approach to description of non-stationary resonance regimes \cite{Man2, Man3, Man4, Man5,Smirnov,Man_Gend}, including coupled pendula dynamics \cite{Man6}. This concept introduces a fundamental non-stationary process of new type which corresponds to maximum possible energy exchange between the oscillators (in particular, pendula) or clusters of oscillators. In essence, the LPTs play in the non-stationary resonance dynamics of finite systems the role similar to that of the NNMs in the stationary theory and in the study of non-stationary yet non-resonant regimes. In terms of LPT, the transition from intense energy exchange between some clusters of oscillators (coherence domains), in particular, weakly coupled pendula, to energy localization in the initially excited cluster (oscillator) can also be predicted \cite{Aubry, Sepul,Man7}. 

However, all existing analytical results in the non-stationary resonance dynamics of finite-dimensional systems relate to some asymptotic limits which are either quasi-linear systems or separated oscillators (pendula) \cite{Man3,Sepul}. In this study we remove these restrictions and admit arbitrary oscillation amplitude in the framework of LPT concept.  Assuming weak coupling, only the closeness to inter-pendula resonance is required. It is shown that such an extension is crucial for revealing the nature of large amplitude dynamics and prediction of necessary conditions for the onset of chaotic regimes. The analytical findings are confirmed by numerical  simulations. We construct Poincar\'{e} sections for the starting equations of motion to verify the obtained analytical results concerning with the revealed dynamical transitions and their connection with manifestation of chaotic behavior.

\begin{figure}
\centering
\includegraphics[width=8 cm]{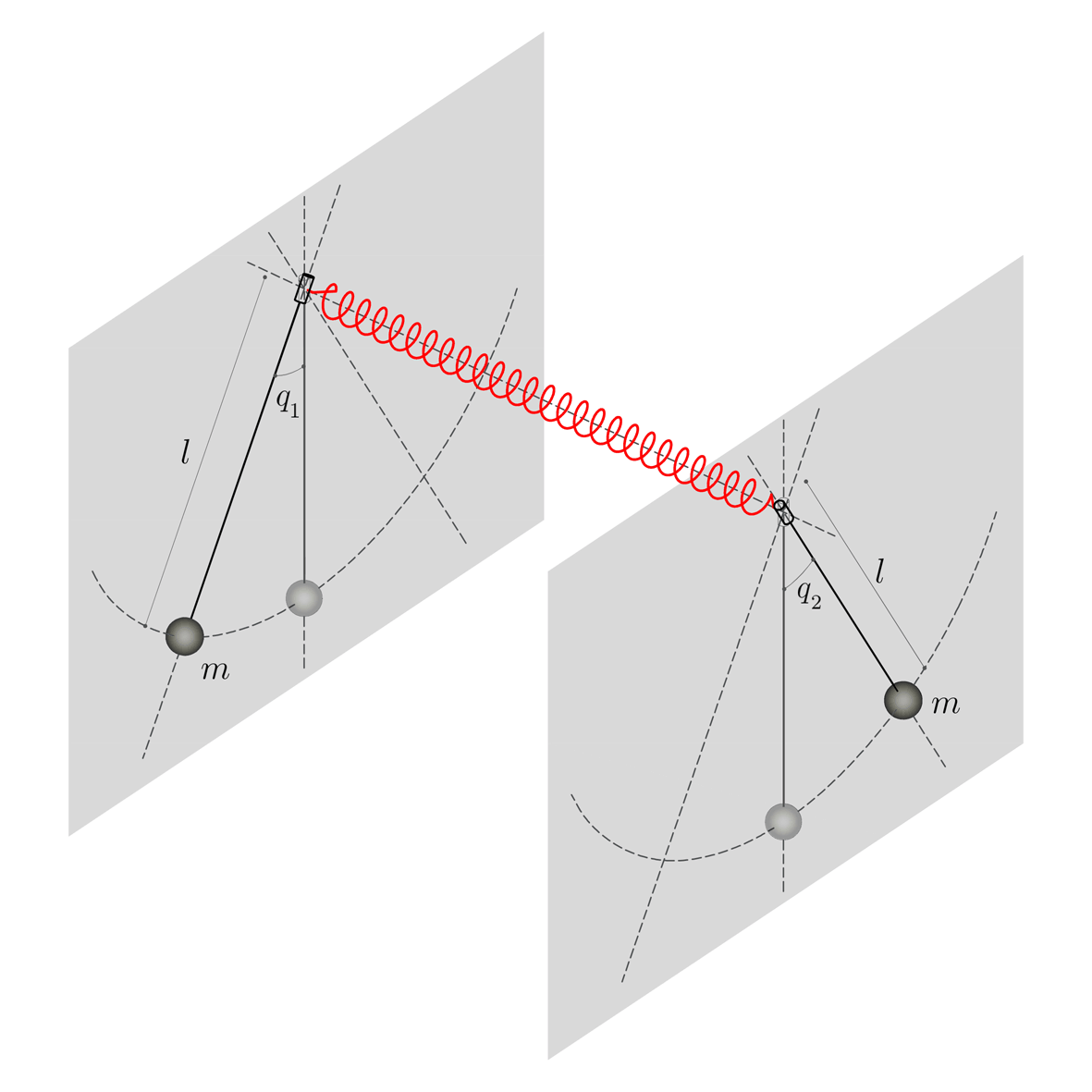}
\caption{Two identical pendula, with unit mass $m$, weakly coupled by a torsional spring.}
\label{fig1} 
\end{figure}
For the sake of clarity we will discuss a mechanical interpretation of the problem in terms of  two weakly coupled pendula undergoing planar motion as sketched in Fig.1. 
 The corresponding dimensionless equations of motion can be written as
\begin{equation}
\frac{d^2 q_j}{d \tau_0^2}+\sin q_j+\varepsilon\beta(q_j-q_{3-j})=0,\,\,\,\,j=1,2
\label{eq_1}
\end{equation}
where $q_j$ is the angular coordinate of the $j$-th pendulum, $\tau_0=\omega_0 t$, $\omega_0=\sqrt{g/l}$ is its linear natural frequency, $g$ is the gravitational acceleration, $\varepsilon \ll 1$ and $\beta$ is the coupling parameter. As known, these equations describe also a particular case of the two Josephson junctions (two-junction interferometer) \cite{Likharev} as well as of the Frenkel-Kontorova model, having numerous applications in solid state physics and photonics \cite{Catal,Braun}.  

Dealing with pendula oscillations under internal 1:1 resonance conditions, we rewrite \eqref{eq_1} in the form
 \begin{equation}
\frac{d^2 q_j}{d \tau_0^2}+\omega^2 q_j+\varepsilon\beta(q_j-q_{3-j})+\varepsilon \mu (\sin q_j-\omega^2 q_j)=0,\,\,j=1,2
\label{eq_2}
\end{equation}
where $0<\omega \le 1$ (the lower limit will be later on specified) is the resonance oscillation frequency. 
Under internal 1:1 resonance conditions the combination of two terms in the second brackets has to be small (we suppose of order-$\varepsilon$), and $\mu=\varepsilon^{-1}$ is a book keeping parameter. Thus, we assume the closeness to resonance but we don't impose any restriction to the oscillations amplitude and the ensuing resonance frequency.
Passing to complex variables, given by
$
\varphi_j =(\frac{d q_j}{d \tau_0}+i\omega q_j)e^{i\omega \tau_0}$; $\varphi_j^* =(\frac{d q_j}{d \tau_0}-i\omega q_j)e^{-i\omega \tau_0}$, and substituting in \eqref{eq_2}, the two scale expansion $\varphi_j=\varphi_{j,0}(\tau_0,\tau_1)+\varepsilon \varphi_{j,1}(\tau_0,\tau_1)+\ldots$, 
in which $\tau_1=\varepsilon\tau_0$ is slow time scale, then taking into account that $\frac{d\cdot}{d\tau_0}=\frac{\partial \cdot}{\partial \tau_0}+\varepsilon \frac{\partial \cdot}{\partial \tau_1}$ and selecting the terms of order $\varepsilon^0$ and $\varepsilon^1$, we get 
 \begin{widetext}
\begin{eqnarray}
 & &\frac{\partial\varphi_{j,0}}{\partial \tau_0}=0 \label{eq_3sm}\\
& &\frac{\partial\varphi_{j,1}}{\partial \tau_0}+ \frac{\partial\varphi_{j,0}}{\partial \tau_1}-\frac{i\beta}{\omega}(\varphi_{j,0}-\varphi_{j,0}^*e^{-2i\omega\tau_0}-
 \varphi_{3-j,0}-\varphi_{3-j,0}^*e^{-2i\omega \tau_0}) +\mu\left[e^{-i\omega\tau_0}\sin \left(\frac{\varphi_{j,0}e^{i\omega\tau_0}-\varphi_{j,0}^*e^{-i\omega\tau_0}}{2i\omega}\right)\right. \nonumber \\
& &\left. -\frac{\omega^2}{2i\omega}\left( \varphi_{j,0}-\varphi_{j,0}^*e^{-2i\omega\tau_0}\right)\right]=0, \quad j=1,2
\label{eq_4sm}
\end{eqnarray}
\end{widetext}
The partial differential equations  \eqref{eq_3sm} show that  $\varphi_{j,0}=\varphi_{j,0}(\tau_1)$ therefore equations \eqref{eq_4sm} can be considered as ordinary differential equations with respect to fast time scale $\tau_0$. While integrating them, the conditions of absence of secular terms lead to the following equation 
%
%
\begin{widetext}
\begin{equation}
\frac{d \varphi_{j,0}}{d \tau_1}-\frac{i\beta}{\omega}(\varphi_{j,0}-\varphi_{3-j,0})+\mu\left[-\frac{i}{2\omega}\varphi_{j,0}+\sum_{n=1}^{\infty}(-1)^n\frac{ \alpha_n}{\omega^{2n+1}} \left( \frac{|\varphi_{j,0}|^2}{8}\right)^n\varphi_{j,0}+\frac{i\omega}{2}\varphi_{j,0}\right]=0, \quad j=1,2
\label{eq_5sm}
\end{equation}
 \end{widetext}
 in which $\alpha_n=1/(2a_n)$ and $a_n$ is obtained by the following recurrence relations $a_n=b_n a_{n-1}$, with $a_0=1$, and $b_n=b_{n-1}+n$, with $b_0=0$.
Equations  \eqref{eq_5sm}  represent the main asymptotic approximation in slow time scale with respect to $\varphi_{j,0}$. Contrary to original system (1) which possesses only one integral, system \eqref{eq_5sm} admits two integrals of motion:
\begin{widetext}
 \begin{eqnarray}
H&=&\left[\frac{i\beta}{\omega}+\mu\left(\frac{i\omega}{2}-\frac{i}{2\omega}\right)\right]\sum_{j=1}^{2}|\varphi_{j,0}|^2-\frac{i\beta}{\omega}(\varphi_{1,0}^*\varphi_{2,0}+\varphi_{2,0}^*\varphi_{1,0})+\sum_{j=1}^{2}\sum_{n=1}^{\infty}(-1)^n\frac{i\alpha_n}{(n+1)\omega^{2n+1}}\left(|\varphi_{j,0}|^2\right)^{n+1}
\label{eq_6sm} \\
N&=&\sum_{j=1}^{2}|\varphi_{j,0}|^2
\label{eq_7sm}
\end{eqnarray}  
\end{widetext}
Taking into account the second integral  \eqref{eq_7sm}, 
we set $\varphi_{1,0}=\sqrt{N}\cos\theta(\tau_1) e^{i\delta_1(\tau_1)}$, $\varphi_{2,0}=\sqrt{N}\sin\theta(\tau_1) e^{i\delta_2(\tau_1)}$; then, the first integral  \eqref{eq_6sm}
, after series summation, takes the following form
\begin{widetext}
\begin{equation}
\hspace{-0.1cm}H\hspace{-0.1cm}=\hspace{-0.05cm}\frac{\beta}{\omega}\cos\Delta\sin2\theta 
\hspace{-0.05cm}+\frac{\mu}{2N\omega}\big\{\hspace{-0.05cm}N\hspace{-0.1cm}-\hspace{-0.05cm}8\omega^2\hspace{-0.1cm}+\hspace{-0.05cm}4\omega^2\hspace{-0.1cm}\left[J_0(k_2\cos\theta)\hspace{-0.1cm}+\hspace{-0.1cm}J_0(k_2\sin\theta)\right]\},
\label{eq_9}
\end{equation}
 \end{widetext}
where $\Delta=\delta_2-\delta_1$ and  $k_2=\sqrt{N}/\omega$. Corresponding equations of motion can be written as follows
\begin{widetext} 
\begin{eqnarray}
\dot \theta&=&-\frac{\beta}{\omega}\sin\Delta \label{eq_10}\\
\dot\Delta\sin2\theta&=&-\frac{2\beta}{\omega}\cos 2\theta\cos\Delta-\frac{2 \mu}{\sqrt{N}}  \left[ J_1\left(k_2\cos \theta 
   \right)\sin \theta -  J_1\left( k_2 \sin \theta 
 \right)\cos \theta\right],
%
\label{eq_11}
\end{eqnarray}
\end{widetext} 
where $J_0()$, $J_1()$ are Bessel functions of the first kind and the derivative with respect to $\tau_1$ is considered. This system describes the slow dynamics of weakly coupled pendula for arbitrary initial conditions.
\begin{figure}
\centering
\includegraphics[width=8.5 cm]{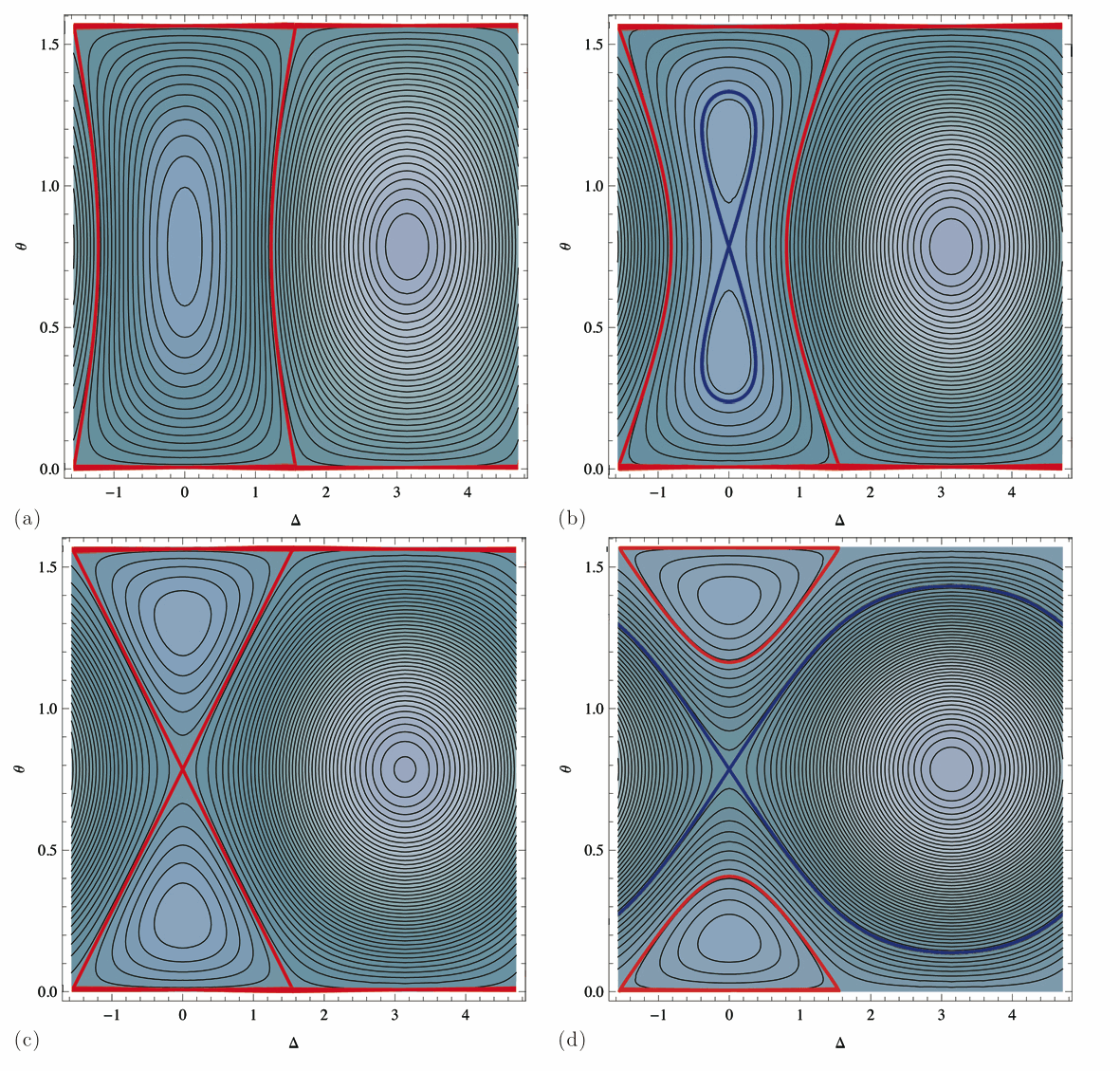}
\caption{Evolution of the $\Delta$-$\theta$ phase portrait for $\omega=0.65$ and maximum angles $q_{j,max}\simeq3\pi/4$ for decreasing parameter $\varepsilon$; LPTs (red), separatrix (blue). a) Before first transition, $\varepsilon=0.2$; b) after first transition, $\varepsilon=0.1$; c) second transition, $\varepsilon=0.0695$; d) after second transition, $\varepsilon=0.05$. }
\label{fig2} 
\end{figure}
Having obtained the integral \eqref{eq_9} and equations of motions \eqref{eq_10}-\eqref{eq_11}  we have the possibility to perform the analytical study of the considered problem. We begin by analyzing the phase portrait described by equation $H$=const.. It is convenient to present the evolution of the $\Delta$-$\theta$ phase plane with changing the parameter $\varepsilon$, characterising interpendulum coupling,  for a given value of resonance frequency that determines also the maximum value $q_{j,max}$ of pendulum oscillation angle $q_j$. In Fig.2 the phase portrait corresponding to $\omega=0.65$ ($q_{j,max}\simeq3\pi/4$) and $\beta=1.0$ is shown.

There exist two dynamical transitions, relating stationary and highly nonstationary dynamics,
that are manifested by changing the parameter $\varepsilon$; the phase portraits shown in Fig.2 highlight the topological changes associated with these transitions. 
The  stationary points in Fig.2 correspond to NNMs of the considered system; the closed phase trajectories surrounding them imply weak energy exchange between the pendula. The Limiting Phase Trajectory encircling all trajectories corresponds to full energy exchange between the pendula and the trajectories close to it describe intense energy exchange. The first dynamical transition occurs due to instability and bifurcation of the in-phase NNM (Fig.2b). It leads to appearance of two additional stable in-phase (yet asymmetric) NNMs, and the homoclinic separatrix encircles them. This is a local transformation of the phase portrait which strongly influences the stationary dynamics of the pendula. However, there is not yet any qualitative change in its highly non-stationary dynamics because complete energy exchange between the pendula can still be possible. Only the second dynamic transition (Fig.2c), caused by global transformation of the phase portrait after coalescence of LPT and homoclinic separatrix leads to a drastic change of non-stationary dynamics which can no longer give rise to full energy exchange between pendula and therefore is characterized by predominant energy localization in the initially excited pendulum. Moreover, after this transition the homoclinic orbit turns into a heteroclinic one (Fig.2d). 

The analytical conditions for both transitions are reported below and are confirmed by the numerical solution of equations \eqref{eq_1}. 
For the first transition prediction we resort to the solution of equations  \eqref{eq_10}-\eqref{eq_11} in the vicinity of the stationary point $\Delta=0$, $\theta=\pi/4$ (in-phase NNM) as $\Delta=\Delta_1$, $\theta=\pi/4+\theta_1$. Assuming that $\Delta_1$ and $\theta_1$ are small perturbations, the solution can be determined by means of the linearized version of equations  \eqref{eq_10}-\eqref{eq_11}:
\begin{eqnarray}
\dot \theta_1&=&-\frac{\beta}{\omega}\Delta_1 \label{eq_7asm}\\
\dot\Delta_1&=&\left[\frac{4\beta}{\omega}+\frac{\mu}{\omega}J_0\left(\frac{k_2}{\sqrt{2}}\right)-2\mu\sqrt{\frac{2}{N}} J_1\left(\frac{k_2}{\sqrt{2}}\right) \right. \nonumber \\
&-&\left.\frac{\mu}{\omega}J_2\left(\frac{k_2}{\sqrt{2}}\right)\right]\theta_1
\label{eq_7bsm}
\end{eqnarray}
From the system \eqref{eq_7asm}-\eqref{eq_7bsm} it is seen that instability of in-phase NNM occurs when the coefficient of $\theta_1$ is equal to zero. The latter condition leads to the following expression for the first transition 
\begin{equation}
\varepsilon=\frac{1}{4\beta}\left[-J_0\left(\frac{k_2}{\sqrt{2}}\right)+2\omega\sqrt{\frac{2}{N}} J_1\left(\frac{k_2}{\sqrt{2}}\right)+J_2\left(\frac{k_2}{\sqrt{2}}\right)\right]
\label{eq_14}
\end{equation}   
after which the phase portrait becomes qualitatively alike the one shown in Figure 2b.
To reveal the condition of the second transition we derive the equation describing LPTs  by considering that they possess the point $\theta=0$ or $\theta=\pi/2$, thus the corresponding Hamilton function can be written as $H_{LPT}=H-C$, in which $C$ is the constant given by $H$ for $\theta=0$. From $H_{LPT}$ and the equations of motion \eqref{eq_10}-\eqref{eq_11} the second order nonlinear differential equation for $\theta$, which is valid for LPTs only, is obtained. The corresponding $(\theta,\dot \theta)$ phase portraits are presented in Fig.3b for several initial conditions. While the phase trajectories out of the separatrix correspond to full energy exchange between pendula, those inside the separatrix correspond to maximum possible energy exchange in the condition of energy localization in the excited pendulum.  If the threshold of first transition is given by \eqref{eq_14} (curve I in Fig.3a), the condition of second transition can be found by taking into account that its occurrence implies that LPT possesses the unstable stationary point $\Delta=0$, $\theta=\pi/4$, therefore we get the sought second transition (curve II in Fig.3a), namely
\begin{equation}
\varepsilon=\frac{2\omega^2}{N}\left[1+J_0\left(k_2\right)-2J_0\left(\frac{k_2}{\sqrt{2}}\right)\right]
 \label{eq_16}
\end{equation}   
\begin{figure*}
\centering
\includegraphics[width=0.75 \textwidth]{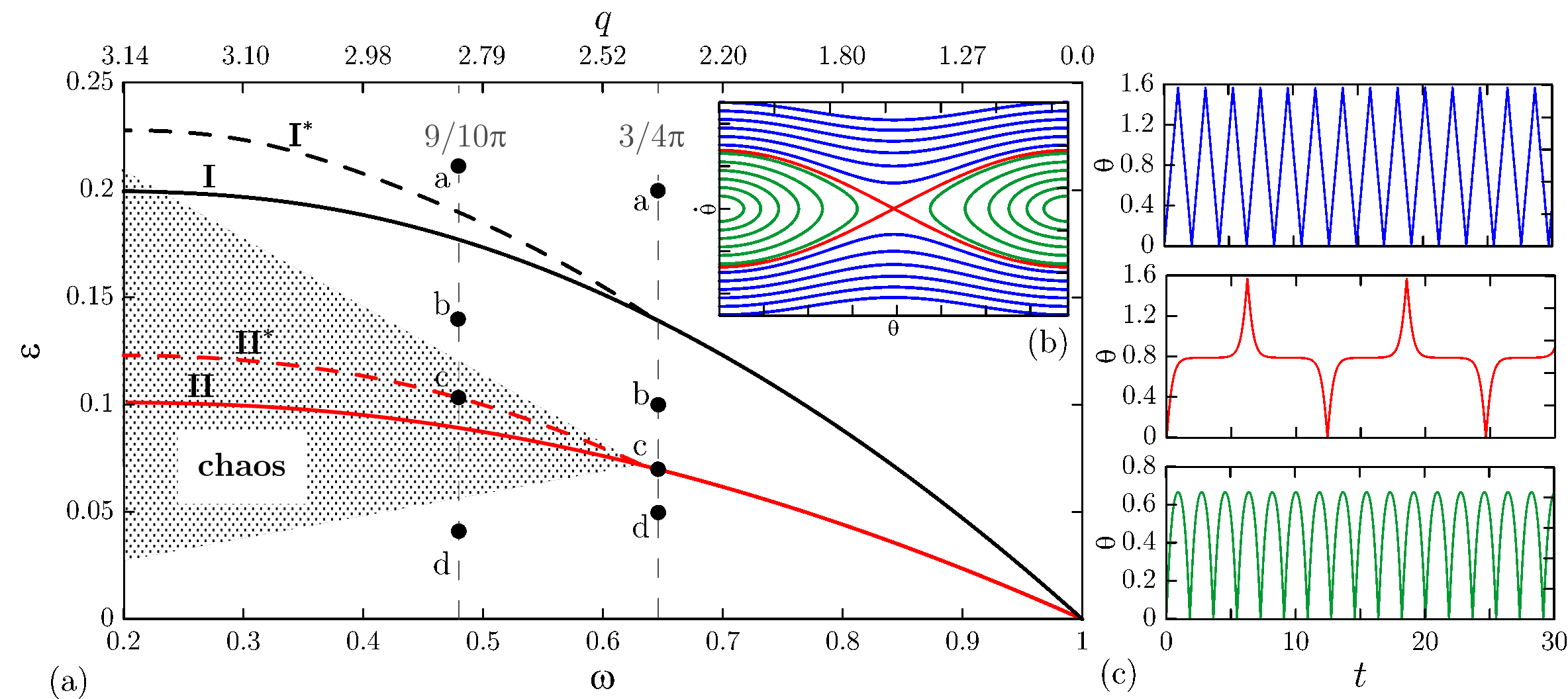}
\caption{Evolution of the dynamic transitions in the ($\omega-\varepsilon$) parameter space. The upper horizontal axis shows the maximum angle $q$ for corresponding $\omega$ in the lower axis. a) first and second dynamic transitions, analytical prediction (I,II) and numerical observation (I*,II*); b) LPTs phase plane; c) temporal behavior of $\theta$ corresponding to the trajectories shown in b).}
\label{fig3} 
\end{figure*}
It is worth emphasizing that the described scenario is observed not only for small angles (quasi-linear case) but also for values of the angles close to $\pi$. However, the threshold values of the parameters corresponding to both transitions change strongly with the resonance frequency, as shown in Fig.3a. 
The perfect agreement between analytical prediction of curves $I,II$ and their numerical counterparts $I^*,II^*$ is observed for $0<q\lesssim3\pi/4$ ($0<\omega\lesssim 0.65$). 
The top horizontal axis labels refer to the amplitude $q$ corresponding to the values of resonance frequency $\omega$ reported in the bottom horizontal axis. It can be seen that the lowest value of the latter resonance frequency is equal to $\omega=0.2$ ($q_{max}=3.14$). Moreover both boundaries shown in Figure 2a refer to parameter $\varepsilon$ ranging from $0$ to $0.2$, in agreement with our initial assumption concerning with the smallness of the sum in the second bracket in equation \eqref{eq_2}. The quantitative difference for larger angles (which reaches at most 10\% and 20\% for curve $I$ and $II$, respectively) can be reduced by considering next order approximation in the multiple scale expansion procedure.
The structure of the phase plane depicted in Fig.3b allows to predict the temporal behavior of the angle variable $\theta$. The trajectories situated far from the separatrix correspond to almost straight lines. However, due to the restriction $0\leqslant\theta\leqslant\pi/2$, they become saw-tooth type functions. The analytical solution of the problem in terms of non-smooth functions can be obtained after change of temporal variable through the procedure proposed in \cite{Pil1, Pil2} and used for the study of non-stationary resonance processes in \cite{Man2, Man4, Man5}.
As for phase trajectories located inside the separatrix, they correspond to localized LPTs and can be easily found after linearization of the second order equation for LPTs in the vicinity of $\theta=\dot\theta=0$. 
By substituting $\theta=0$ in \eqref{eq_9} we get $H=C$, where $C$ is the constant 
$$
C=\frac{\mu  \left(4 \omega ^2 J_0\left(k_2\right)+N-4 \omega
   ^2\right)}{2 N \omega }
$$   
Thus, $H_{LPT}=H-C$ and the corresponding expression for $\cos\Delta$ reads
\begin{widetext}
\begin{equation}
\cos\Delta=\frac{\mu  \omega ^2 \csc (\theta ) \sec (\theta
   ) \left(-J_0\left(\sin (\theta )
   k_2\right)-J_0\left(\cos (\theta )
   k_2\right)+J_0\left(k_2\right)+1\right)}{N
   \beta }     \label{eq_9sm}
\end{equation}
\end{widetext}
Expression \eqref{eq_9sm} can be substituted into the first derivative of \eqref{eq_7asm} given by 
\begin{equation}
\ddot \theta=-\frac{\beta}{\omega}\cos\Delta \dot \Delta
  \label{eq_10sm}
\end{equation}
and $\dot \Delta$ is obtained from \eqref{eq_7sm}, leading to the sought second order differential equation
\begin{widetext}
\begin{eqnarray}
\ddot \theta&=&\frac{4 \mu ^2 \omega ^{3/2}}{N^2} \csc ^2 2 \theta 
   \left[-J_0\left(\sin \theta  k_2\right)-J_0\left(\cos
   \theta  k_2\right)+J_0\left(k_2\right)+1\right] \nonumber \\
  && \left[-\sqrt{N} \cos \theta  J_1\left(k_2 \sqrt{\omega }  \sin
   \theta \right)+\sqrt{N} \sin \theta 
   J_1\left(k_2 \sqrt{\omega }  \cos \theta \right)+2
   \sqrt{\omega } J_0\left(k_2\right) \cot 2 \theta \right. \nonumber \\
   && \left. -2 \sqrt{\omega } \cot (2 \theta ) J_0\left(k_2\cos \theta 
   \right)-2 \sqrt{\omega } \cot 2 \theta  J_0\left(k_2\sin
   \theta  \right)+2 \sqrt{\omega } \cot 2 \theta
   \right]    \label{eq_11sm}
\end{eqnarray}
\end{widetext}
Equation \eqref{eq_11sm} allows to construct the LPTs phase plane shown in Figure 3b.\\
It is worth emphasizing that Fig.3a clarifies the relation among the obtained results and conventional approximations used for description of coupled pendula dynamics: the quasi-linear approaches can be applied only in the right part of the parametric plane ($0<q\lesssim\pi/4$) whereas the independent pendula approximation holds only for the bottom part of the parametric plane ($0<\varepsilon\lesssim0.03$). On the contrary, the proposed approach based on resonance asymptotic turns out to be valid for description of regular motion in all parametric plane ($\omega$-$\varepsilon$). 

\begin{figure}[bh]
\centering
\includegraphics[width=8.5 cm]{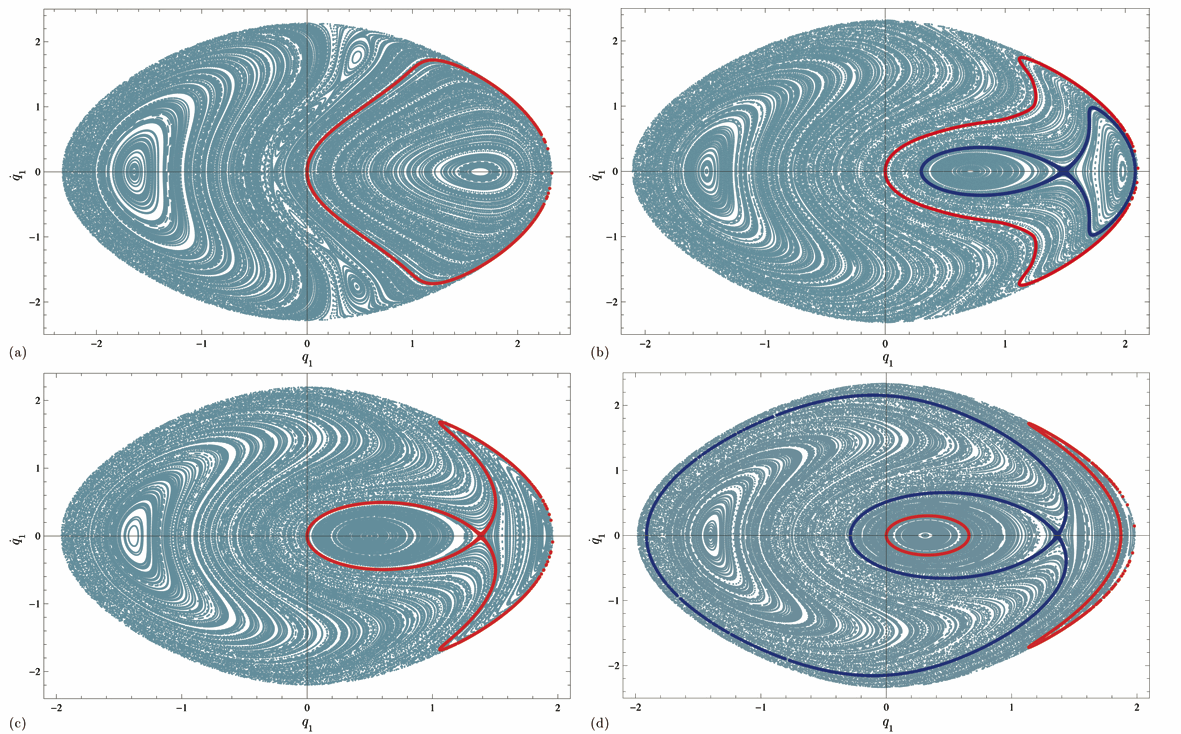}
\caption{Poincar\'{e} sections for $\omega=0.65$ and angles $q_{j,max}\simeq3\pi/4$ for decreasing parameter $\varepsilon$; LPTs (red), separatrix (blue). a) Before first transition, $\varepsilon=0.2$; b) after first transition, $\varepsilon=0.1$; c) second transition, $\varepsilon=0.0695$; d) after second transition, $\varepsilon=0.05$.}
\label{fig3} 
\end{figure}

So, we have found the dependence on the resonance frequency (or initial angle) of the thresholds corresponding to both dynamical transitions. These relations allow to single out in parametric space the domains of regular motion of the pendula. It has been shown that all regular motions revealed in main asymptotic approximation are observed also by integrating the starting equations of motion \eqref{eq_1}. Moreover, the analytical predictions of both dynamic transitions are well confirmed as well. However, it must be underlined that, contrary to the asymptotic approximation,  the initial system \eqref{eq_1} is not integrable. Therefore it is of interest to clarify the onset of chaotic behavior and the role played by LPTs in the general behavior of the pendula. Towards this goal, Poincar\'{e} sections constructed on the basis of the starting equation of motions \eqref{eq_1} are reported in this section. In Fig.4 and Fig.5 Poincar\'{e} sections are shown referring to  maximum angles $q_{j,max}\simeq3\pi/4$ and  $q_{j,max}\simeq9\pi/10$, respectively. The four sections correspond to different dynamic regimes (see Fig.2) for decreasing values of $\varepsilon$, according to the points highlighted in Fig.3a, Fig.4a and Fig.5a refer to the dynamics before the first transition where the LPTs (red curve) encircling the in-phase NNM are also depicted. Fig.4b and Fig.5b refer to the case in-between the two transitions,  where the new stationary states born as a result of instability of in-phase NNM can be seen; moreover the associated homoclinic separatrix encircles the corresponding stationary points. Fig.4c and Fig.5c reflect the conditions at second transition, where LPT becomes separatrix; as indicated in Fig.3a, manifestation of chaotic behavior can be observed in the vicinity of second dynamic transition for large enough angles ($q_{j,max}\simeq9\pi/10$, Fig.5c). In Fig.4d and Fig.5d the localized LPTs as well as the heteroclinic separatrix (blue curve) are well seen; it is worth noticing that, for small enough $\varepsilon$, the motion remains regular in all phase-space (see also Fig.3a).   

\begin{figure}
\centering
\includegraphics[width=8.5 cm]{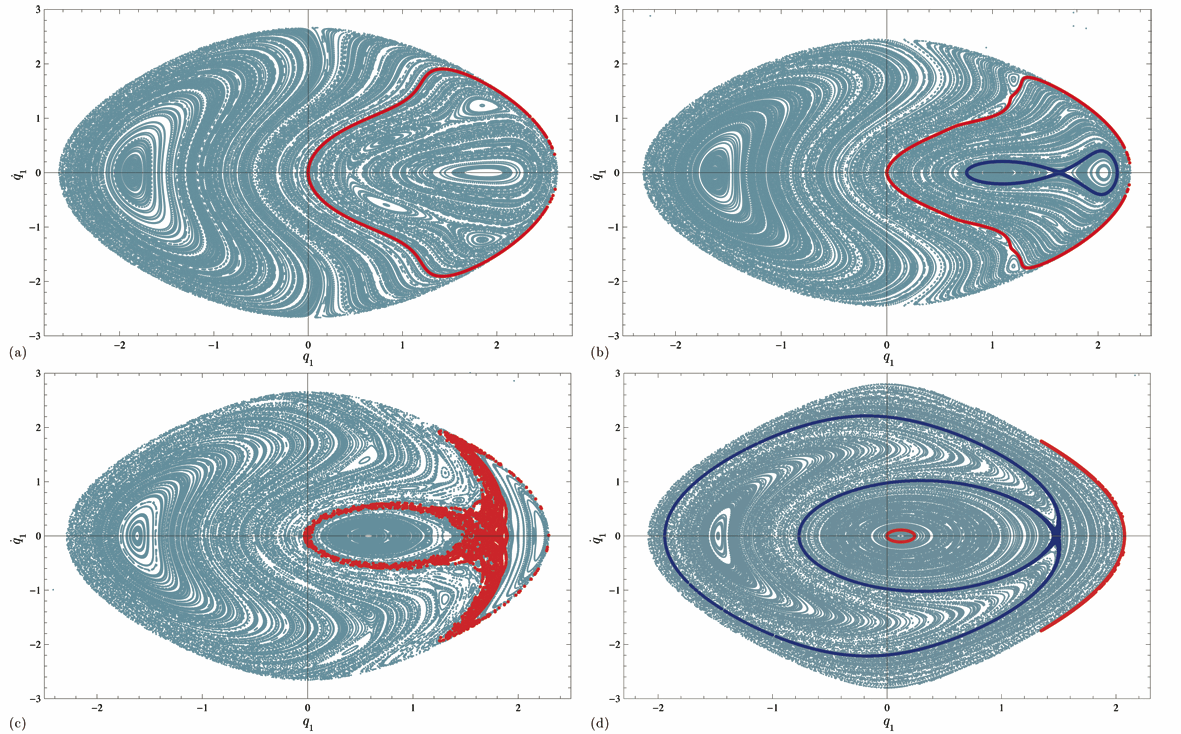}
\caption{Poincar\'{e} sections for $\omega=0.482$ and angles $q_{j,max}\simeq9\pi/10$ for decreasing parameter $\varepsilon$; LPTs (red), separatrix (blue). a) Before first transition, $\varepsilon=0.225$; b) after first transition, $\varepsilon=0.145$; c) second transition, $\varepsilon=0.104$; d) after second transition, $\varepsilon=0.03$.}
\label{fig4} 
\end{figure}
Examples of the pendula oscillations temporal evolution corresponding to the parameters considered Figures 4,5 are reported in the following Figures 6,7. The pendula response reported is obtained by direct numerical integration of the initial dimensionless equations of motion \eqref{eq_1}. More specifically, Figure 6 refers to the case $\omega=0.65$ and the maximum angles $q_j$ are $3/4\pi$ for which only regular behavior was observed (see Fig. 3a). Differently, in Figure 7 the case $\omega=0.482$ and the maximum angles $q_j$ are $9/10\pi$ is considered. In agreement with the findings shown in Fig 3a, chaotic behavior can be observed in Figure 4c in the vicinity of second tansition.
\begin{figure}[htb]
\centering
\includegraphics[width=8.5 cm]{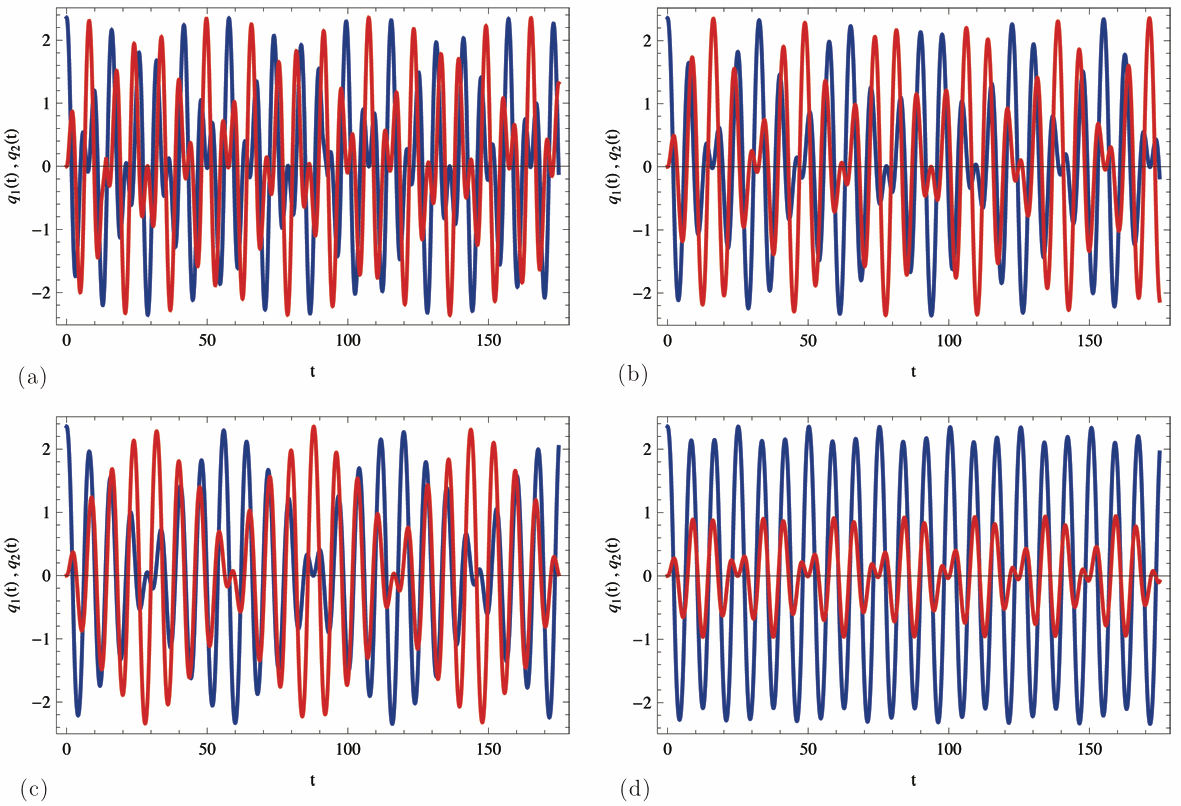}
\caption{Time histories of the two pendula oscillations for $q_1(0)=3\pi/4$ and decreasing coupling parameter $\epsilon$: $q_1(t)$ (blue), $q_2(t)$ (red). a) Before first transition, $\epsilon=0.2$; b) after first transition, $\epsilon=0.1$; c) second transition, $\epsilon=0.0695$; d) after second transition, $\epsilon=0.05$.}
\label{fig2sm} 
\end{figure}
\begin{figure}[h]
\centering
\includegraphics[width=8.5 cm]{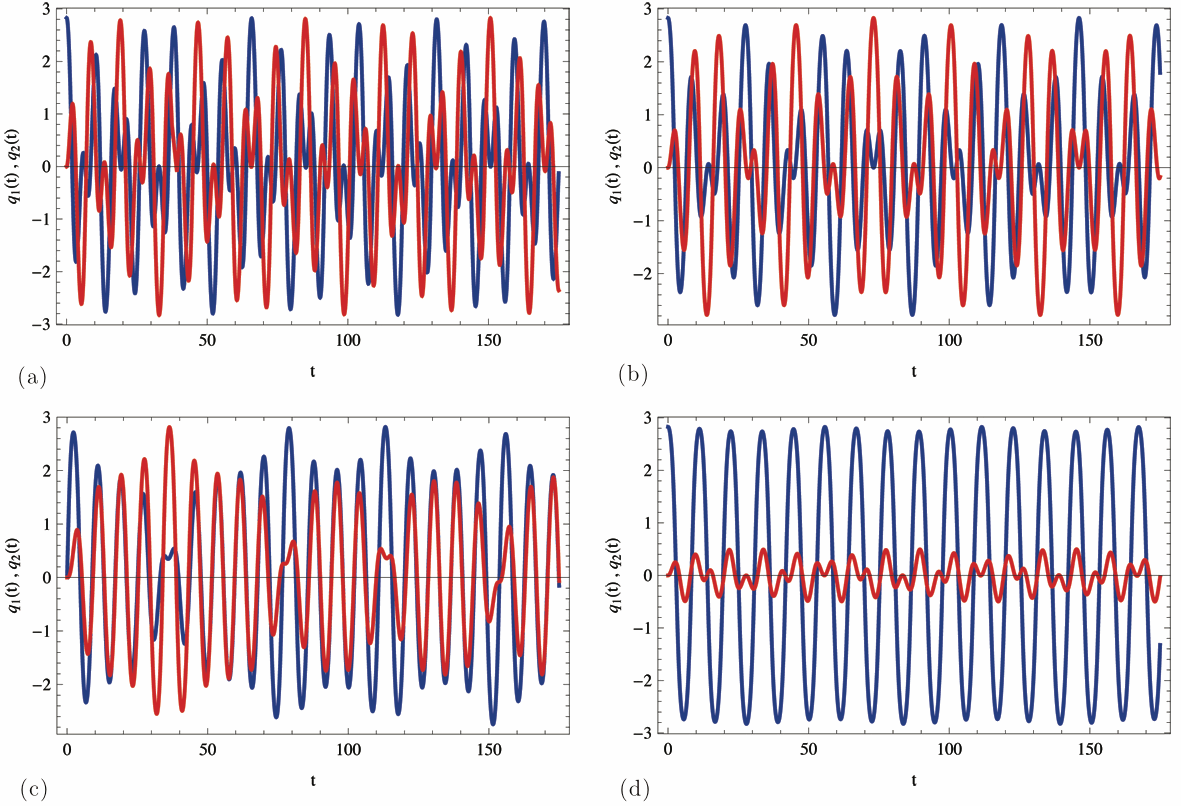}
\caption{Time histories of the two pendula oscillations for $q_1(0)=9\pi/10$ and decreasing coupling parameter $\epsilon$: $q_1(t)$ (blue), $q_2(t)$ (red). a) Before first transition, $\epsilon=0.225$; b) after first transition, $\epsilon=0.145$; c) second transition, $\epsilon=0.104$; d) after second transition, $\epsilon=0.03$.}
\label{fig3sm} 
\end{figure}

While only regular motion is observed for angles less than 135$^\circ$, for larger angles signs of chaotization appear in the vicinity of the second dynamic transition triggered by heteroclinic chaos, as anticipated in the discussion of Fig.3a.
As the maximum oscillation angle grows, the chaotic region in the parameter space increases on both sides of the second transition threshold. Then, for angles greater than 170$^\circ$, this region approaches the first dynamic transition threshold where homoclinic chaos occurs.
The absence of chaotization for maximum oscillation angles less than 135$^\circ$ means that the system is close to integrable separated pendula. For larger angles the system is far enough from being integrable and chaos appears. Such behavior is caused by interaction of dynamic separatrix, coinciding with LPT in the conditions of second dynamic transition, and the conventional pendulum separatrix.

Summarizing, the analytical description of highly non-stationary resonance processes in a system of weakly coupled pendula without any restrictions on the amplitude of oscillations was presented for the first time. It is shown that such processes can be adequately described by LPTs corresponding to maximum possible energy exchange between pendula. These regimes encircle the domains of regular motion which are determined for all initial angles in oscillation dynamic regime. It is also shown that manifestation of chaotic behavior in the considered model is strongly connected with a purely non-stationary dynamic transition.

The obtained results can be applied in the variety of fields where the model of weakly coupled pendula plays a basic role. For example, in the application to Josephson junctions they correspond to a particular case in which both damping and external forces can be taken into account in the next order approximation \cite{Likharev}. Therefore, the revealed regimes with intense inter-pendulum energy exchange and predominant energy localization in one of the two pendula can be experimentally verified and exploited in numerous applications of Josephson junctions.  

\nocite{*}

\begin{thebibliography}{99}
\bibitem{Scott}
A. Scott, {\em Nonlinear Science}, Oxford University Press, New York (2003).
\bibitem{Braun}
O.M. Braun, Y.S. Kivshar, {\em The Frenkel-Kontorova Model}, Springer-Verlag, Berlin Heidelberg (2004).
\bibitem{Likharev}
K.K. Likharev, {\em Dynamics of Josephson Junctions and Circuits}, Gordon and Breach Science Publishers Amsterdam (1986).
\bibitem{Hadley}
P. Hadley, M.R. Beasley, K. Wiensenfeld, Phys. Rev. B {\bf 38}, 8712-8719 (1988).
\bibitem{Braun2}
O.M. Braun, Surface Science {\bf 230}, 262-276 (1990).
\bibitem{Catal}
F. S. Cataliotti,S. Burger, C. Fort, P. Maddaloni, F. Minardi, 
A. Trombettoni, A. Smerzi, M. Inguscio, Science {\bf 293}, 843-846 (2001).
\bibitem{Yaku}
L. V. Yakushevich, A. V. Savin, and L. I. Manevitch, Phys. Rev. E. {\bf 66}, 016614 (2002).
\bibitem{Man1}
L.I. Manevitch, Y.V. Mikhlin, V.N. Pilipchuk, {\em The method of normal oscillations for essentially nonlinear systems}, Nauka, Moscow (1989).
\bibitem{Vak1}
A.F. Vakakis, L.I. Manevitch, Y.V. Mikhlin, V.N. Pilipchuk, A.A. Zevin, {\em Normal Modes and Localization in Nonlinear Systems}, Wiley New York (1996).
\bibitem{Nayfeh}
A.H. Nayfeh, D.T. Mook, {\em Nonlinear oscillations}, Wiley, New York (2008).
\bibitem{Sepul}
J.A. Sepulchre, in {\em Localization and Energy Transfer in Nonlinear Systems}, Ed. L. Velazquez, World Scientific (2003).
\bibitem{Man_A}
A.I. Manevitch, L.I. Manevitch {\em The Mechanics of Nonlinear Systems with Internal Resonances}, Imperial College Press, London (2005).
\bibitem{Man2}
L.I. Manevitch, Arch. Appl. Mech. {\bf 77}, 301-312 (2007).
\bibitem{Man3}
L.I. Manevitch, V.V. Smirnov, Phys. Rev. E {\bf 82}, 036602(1Ð9) (2010). 
\bibitem{Man4}
L.I. Manevitch, A.S. Kovaleva, D.S. Shepelev, Physica D {\bf 240}, 1-12 (2011).
\bibitem{Man5}
L.I. Manevitch, M.A. Kovaleva, V.N. Pilipchuk, Europhys. Lett. {\bf 101}, 50002 (2013).
\bibitem{Smirnov}
V.V. Smirnov, D.S. Shepelev, L.I. Manevitch, Phys. Rev. Lett. {\bf 113}, 135502 (2014).
\bibitem{Man_Gend}
L.I. Manevitch, O. Gendelman, {\em Tractable Models of Solid Mechanics}, Springer-Verlag, Berlin Hidelberg (2011).
\bibitem{Man6}
V.V. Smirnov, L.I. Manevitch, Acoust. Phys. {\bf 57}, 271(2011).
\bibitem{Man7}
L.I. Manevitch, in {Vibro-Impact Dynamics of Ocean Systems, vol. 44 of Lecture Notes in Applied and Computational Mechanics}, Springer, Berlin (2009).
\bibitem{Aubry}
S. Aubry, G. Kopidakis, A.M. Morgante, G.P. Tsironis, Physica B {\bf 296}, 222Ð236 (2001) .
\bibitem{Pil1}
V.N. Pilipchuk,  J. Sound Vib., {\bf 192}(1), 43Ð64 (1996).
\bibitem{Pil2}
V.N. Pilipchuk, {\em Nonlinear Dynamics: Between Linear and Impact
Limits}, Springer Verlag, Berlin (2011).
\end{thebibliography}

\end{document}